\documentclass[]{aastex62}

\newcommand{\kms}{\text{kms}^{-1}}

\newcommand{\kpc}{\text{kpc}}

\newcommand{\Gyr}{\text{Gyr}}
\graphicspath{{./}{figures/}}
\shorttitle{Incomplete vertical phase-mixing}
\shortauthors{Monari et al.}

\begin{document}

\title{Coma Berenices: first evidence for incomplete vertical phase-mixing \\ in local velocity space with RAVE -- confirmed with Gaia DR2}

\correspondingauthor{G. Monari}
\email{gmonari@aip.de}

\author{G. Monari}
\affil{Leibniz-Institut fuer Astrophysik Potsdam (AIP), An der Sternwarte 16, 14482 Potsdam, Germany}

\author{B. Famaey}
\affiliation{Universit\'{e} de Strasbourg, Observatoire astronomique de Strasbourg, CNRS UMR 7550, 11 rue de l'Universit\'{e}, 67000 Strasbourg, France}

\author{I. Minchev}
\affil{Leibniz-Institut fuer Astrophysik Potsdam (AIP), An der Sternwarte 16, 14482 Potsdam, Germany}

\author{T. Antoja}
\affil{Institut de Ci\`{e}ncies del Cosmos, Universitat de Barcelona (IEEC-UB), Mart\'{i} i Franqu\`{e}s 1, E-08028 Barcelona, Spain}

\author{O. Bienaym\'{e}}
\affiliation{Universit\'{e} de Strasbourg, Observatoire astronomique de Strasbourg, CNRS UMR 7550, 11 rue de l'Universit\'{e}, 67000 Strasbourg, France}

\author{B.~K. Gibson}
\affiliation{E.A. Milne Centre for Astrophysics, University of Hull, Hull, HU6 7RX, United Kingdom}

\author{E.~K. Grebel}
\affiliation{Astronomisches Rechen-Institut, Zentrum f\"{u}r Astronomie der Universit\"{a}t Heidelberg, M\"{o}nchhofstrasse 12-14, D-69120 Heidelberg, Germany}

\author{G. Kordopatis}
\affiliation{Laboratoire Lagrange, Universite C\^{o}te d'Azur, Observatoire de la C\^{o}te d'Azur, Boulevard de l'Observatoire, CS 34229, 06304, Nice, France}

\author{P. McMillan}
\affiliation{Lund Observatory, Lund University, Department of Astronomy and Theoretical Physics, Box 43, SE-22100, Lund, Sweden}

\author{J. Navarro}
\affiliation{Department of Physics and Astronomy, University of Victoria, Victoria, BC V8P 5C2, Canada}

\author{Q.~A. Parker}
\affiliation{Department of Physics, Chong Yuet Ming Physics Building, The University of Hong Kong, Pok Fu Lam, Hong Kong}

\author{A.~C. Quillen}
\affiliation{Department of Physics and Astronomy, University of Rochester, Rochester, NY 14627, USA}

\author{W. Reid}
\affiliation{Department of Physics and Astronomy, Macquarie University, NSW 2109, Australia}

\author{G. Seabroke}
\affiliation{Mullard Space Science Laboratory,  Holmbury St. Mary, Dorking, Surrey, UK}

\author{A. Siebert}
\affiliation{Universit\'{e} de Strasbourg, Observatoire astronomique de Strasbourg, CNRS UMR 7550, 11 rue de l'Universit\'{e}, 67000 Strasbourg, France}

\author{M. Steinmetz}
\affil{Leibniz-Institut fuer Astrophysik Potsdam (AIP), An der Sternwarte 16, 14482 Potsdam, Germany}

\author{R.~F.~G. Wyse}
\affil{Department of Physics \& Astronomy, Johns Hopkins University, Baltimore, MD 21218, USA}

\author{T. Zwitter}
\affil{Faculty of Mathematics and Physics, University of Ljubljana, Jadranska 19, 1000 Ljubljana, Slovenia}

\section*{}

Before the publication of the Gaia DR2 \citep{GaiaDR2} we confirmed with RAVE \citep{RAVE} and TGAS \citep{GaiaDR1} an observation recently made with the GALAH survey \citep{GALAH} by \cite{Quillen2018} concerning the Coma Berenices moving group in the Solar neighbourhood, namely that it is only present at negative Galactic latitudes. This allowed us to show that it is coherent in vertical velocity, providing a first evidence for incomplete vertical phase-mixing. We estimated for the first time from dynamical arguments that the moving group must have formed at most $\sim 1.5~\Gyr$ ago, and related this to a pericentric passage of the Sagittarius dwarf satellite galaxy. The present note is a rewritten version of the original arXiv post on this result (arXiv:1804.07767), now also including a confirmation of our finding with Gaia DR2.

Coma Berenices is a low velocity moving group that has been repeatedly identified in analyses of the local stellar kinematics \citep[e.g.,][]{Kushniruk2017}. With $(U,V,W)$ being the heliocentric cartesian velocities, with $U$ pointing towards the Galactic centre, $V$ in the sense of the Galactic rotation, and $W$ towards the Galactic North pole, Coma Berenices is localized in the velocity plane at $(U,V) \sim (-10,-5)~\kms$. A fraction of the moving group stars belong to an open cluster \citep{Odenkirchen1998}. 

\begin{figure*}
 \includegraphics[width=0.32\columnwidth]{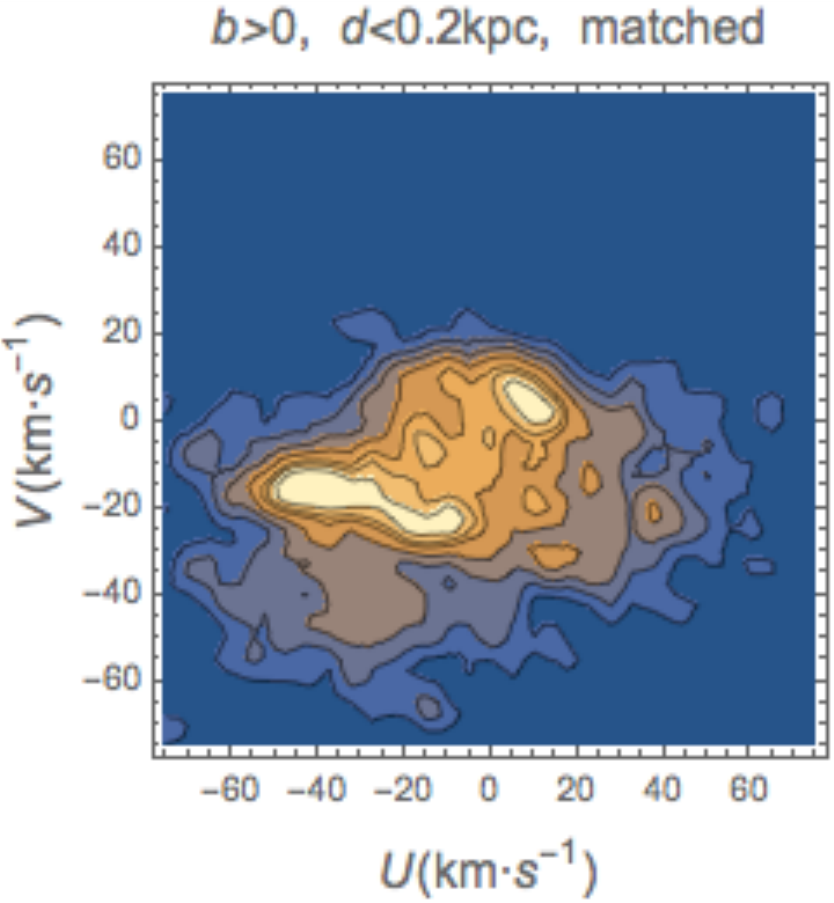}
 \includegraphics[width=0.32\columnwidth]{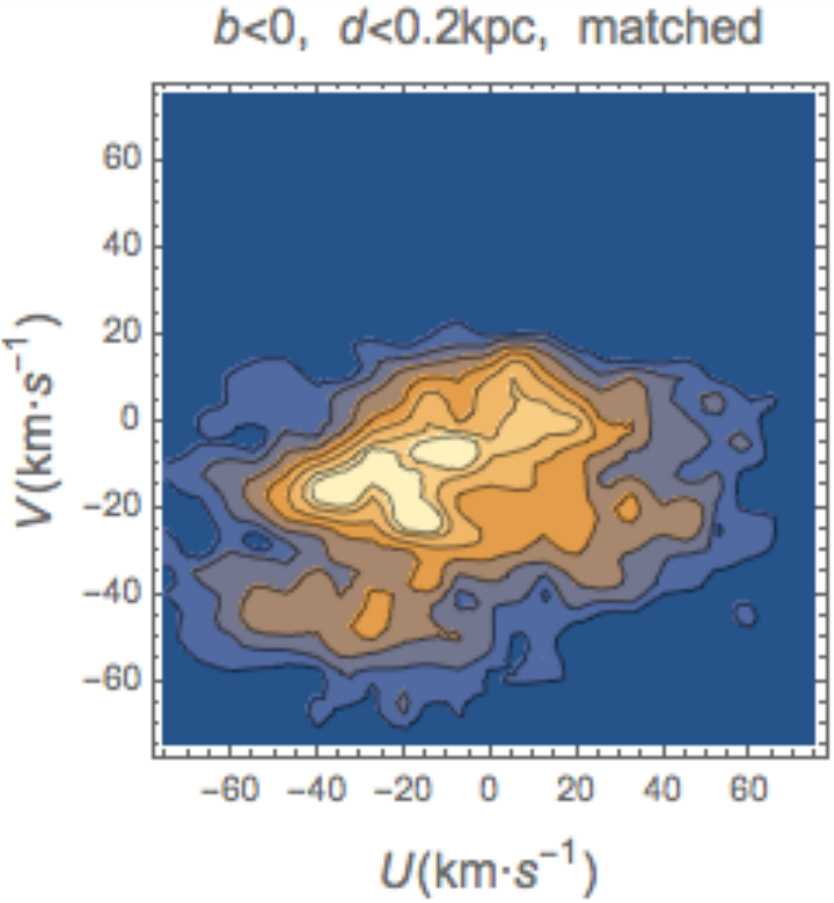}
  \includegraphics[width=0.32\columnwidth]{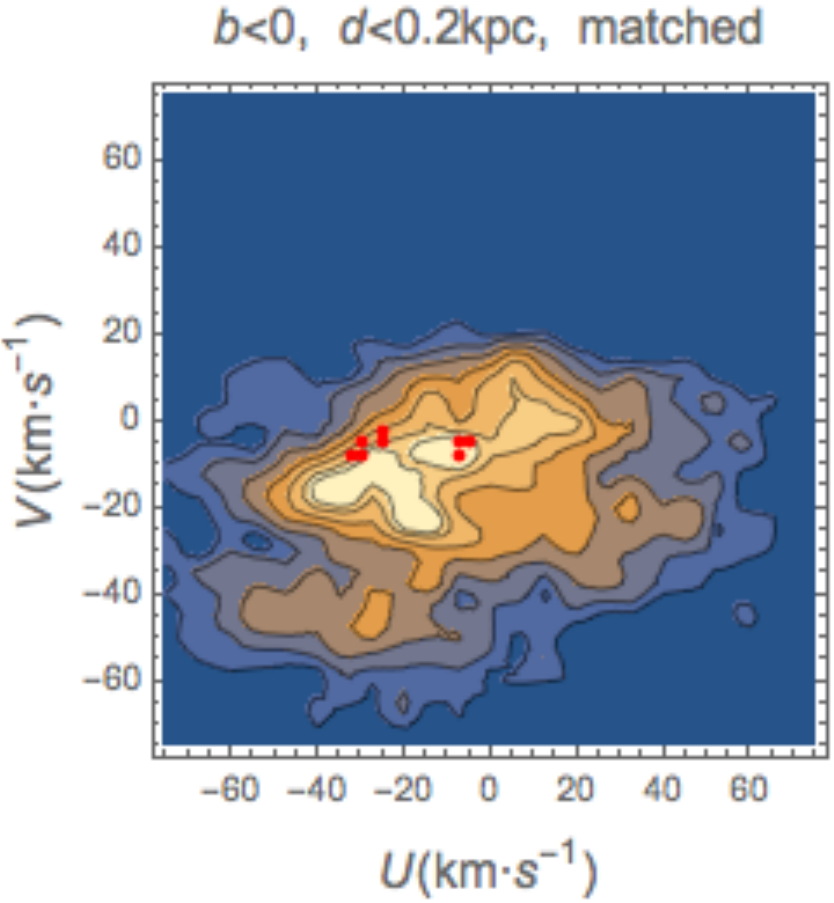}\\
  \includegraphics[width=0.32\columnwidth]{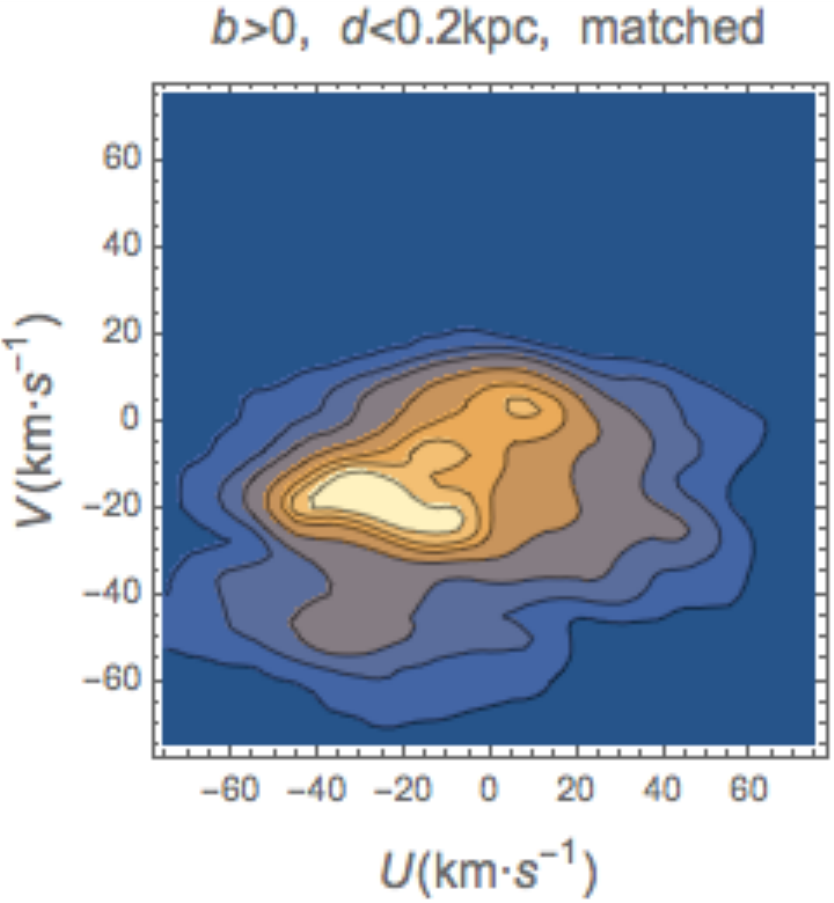}   \includegraphics[width=0.32\columnwidth]{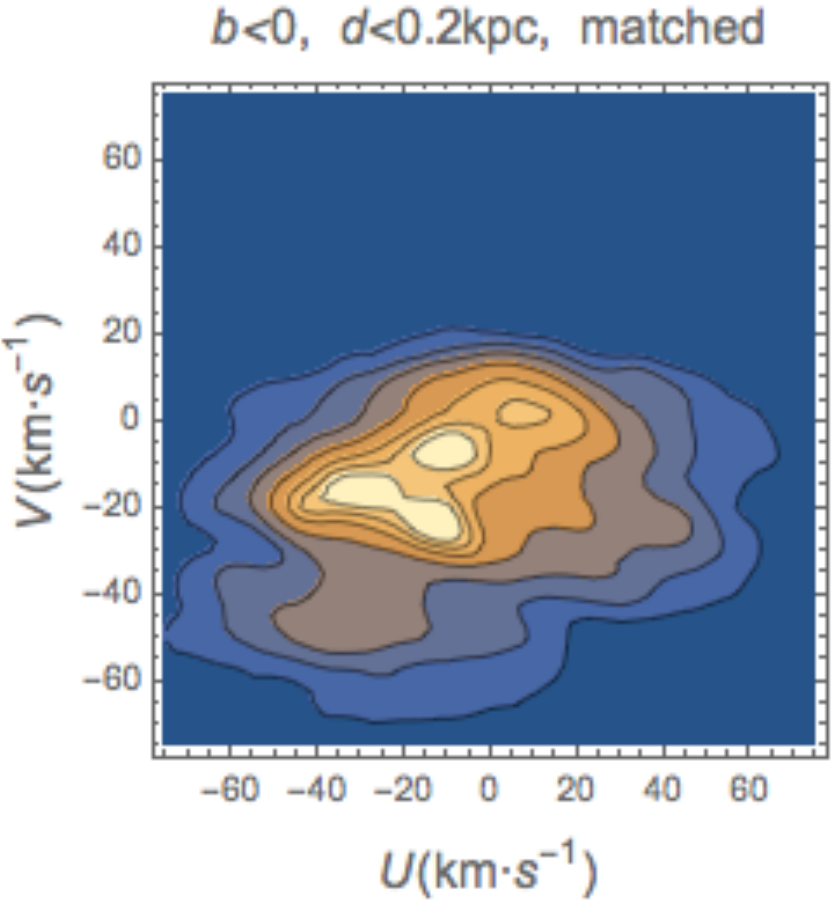}    \includegraphics[width=0.32\columnwidth]{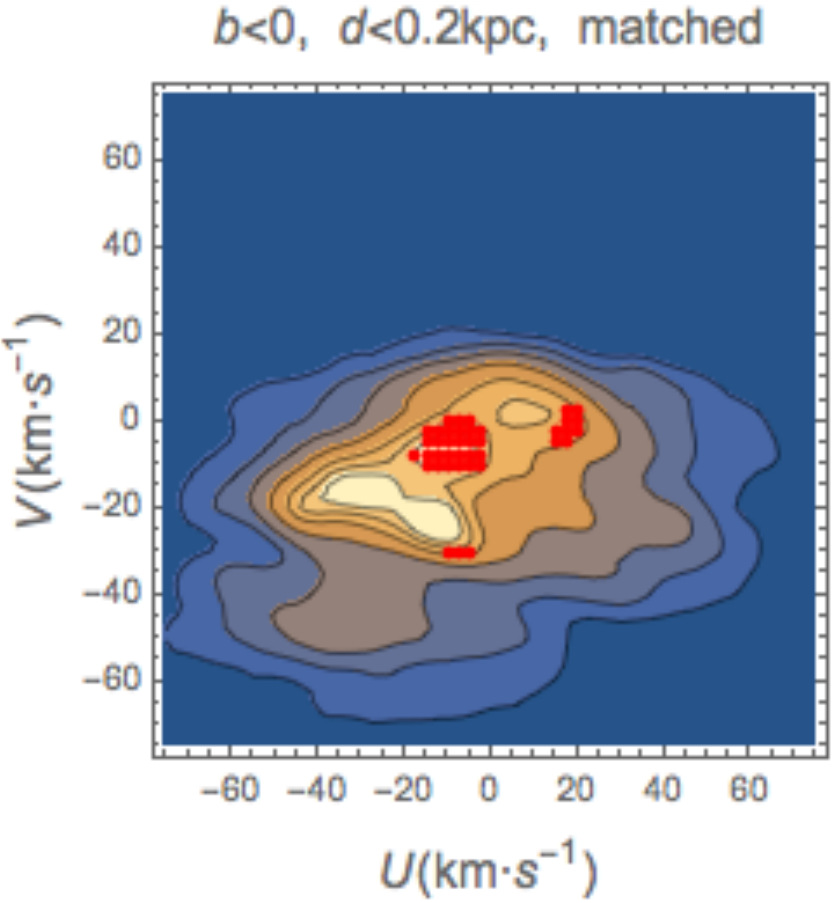}
\caption{Top row: $(U,V)$ velocity distributions for the RAVE DR5 sample (red dots correspond to $2\sigma$ North-South differences). Bottom row: as in the top row but for Gaia DR2 and the red dots correspond to $3\sigma$ differences.}
    \label{fig:vdist}
\end{figure*}

We used stars from the RAVE DR5 \citep{Kunder2017}, with RAVE line-of-sight velocities, TGAS proper motions, and distances $d$ obtained with the RAVE spectrophotometic pipeline using TGAS parallaxes as priors by \cite{McMillan2017}. We selected 73110 stars with converging spectroscopic pipeline, signal-to-noise ratio $SNR>50$, and distance uncertainty $<20\%$. To minimize the effects of the RAVE selection function, we selected stars in the range $l<15^\circ$, $l>230^\circ$, which is common to both samples, and downsized the $b<0^\circ$ sample, to match the smaller number of stars of the $b>0^\circ$ sample, obtaining two sets of 4505 stars each. We then used a statistical bootstrap technique to obtain their density difference, with its variance estimated over 100 resamplings of the sets of stars. In Fig.~\ref{fig:vdist} we show with red dots the regions of the 3 most inner contours of the velocity distribution where the velocity distribution at $b>0^\circ$ is underdense with respect to $b<0^\circ$, with at least $2\sigma$ significance. The inspection of the $(V,W)$ distribution for $b<0^\circ$ and $d<0.2~\kpc$ 
shows that Coma Berenices is coherent in $W$, with the peak displaced of a few $\kms$ from the peak of the whole distribution. A Kolmogorov-Smirnov test excludes that the two samples are drawn from the same distribution (the $p-$value is $p=1.52565\times10^{-4}$).

From the range of $V$ of Coma Berenices we could estimate a range of guiding radii that compose the moving group, using the formula
\begin{equation}
R_\mathrm{g}^2\Omega\left(R_\mathrm{g}\right) \approx R_0 (V + V_\odot + v_0),
\end{equation}
and solving for $R_\mathrm{g}$. Assuming $R_0=8.34~\kpc$, $v_0=240~\kms$ and $V_\odot=12.24~\kms$ \citep{Reid2014,Schonrich2010} we found the guiding radii between $R\sim 8.4~\kpc$ and $R\sim 8.7~\kpc$.

Coma Berenices is localized only at negative $b$, and it seems unlikely that it is caused by a resonance with the bar or the spiral arms, since their distribution is plane-symmetric. Still being clumped in velocity space, the moving group is likely a structure that did not undergo phase-mixing in the Galactic potential. 

An estimate of the phase-mixing time scale is the inverse of the spread in orbital frequencies of the stars: the circular frequency $\Omega$, the epicyclic frequency $\kappa$, and the vertical frequency $\nu$. We estimate their spread from the spread in guiding radii of Coma Berenices, and the Milky Way potential Model 1 from \cite{BT2008} (properly scaled for $R_0$). This correspond to the mixing time scales 
\begin{equation}
T_{\phi} = 2\pi / \Delta\Omega \sim 5.6~\Gyr, \; T_R = 2\pi / \Delta\kappa \sim 3.7~\Gyr, \; T_z = 2\pi / \Delta\nu \sim 1.5~\Gyr.
\end{equation}
Since Coma Berenices is coherent in $U$, $V$, and $W$, we can take the shortest of the three time scales, i.e. $T_z \sim 1.5~\Gyr$, as an upper limit for the moving group's age. Various models of the orbit of the Sgr dwarf galaxy predict a pericentric passage within the Galactic plane, in the anticenter direction, between 500~Myr and 1.5~Gyr ago \citep{Laporte2017}. 
It is thus tempting to associate the vertically phase-unmixed Coma Berenices group to such a perturbation of the disc which could have left its mark in the solar neighbourhood until today \citep[see also][]{Minchev2009}.

The DR2 of Gaia took place after the original finding in this work (arXiv:1804.07767). Using 353327 stars at $d<0.2~\kpc$ from the Gaia DR2 with line-of-sight velocities and distance estimates using the method described by \cite{ABJ2016}, we then clearly confirmed our result in the bottom line of Fig.~\ref{fig:vdist}, where the red dots are in this case the $3\sigma$ detections of the North-South asymmetry of Coma Berenices.

\bibliographystyle{aasjournal}
\bibliography{Comabib} 

\end{document}